# Mechanism of dynamic reorientation of cortical microtubules due to mechanical stress

*Alexander Muratov[1] and Vladimir A. Baulin[1]*

[1] Departament d'Enginyeria Quimica, Universitat Rovira i Virgili 26 Av. dels Paisos Catalans, 43007 Tarragona, Spain

Corresponding author: Vladimir Baulin

Corresponding author's address: Departament d'Enginyeria Quimica, Universitat Rovira i Virgili 26 Av. dels Paisos Catalans, 43007 Tarragona, Spain

Corresponding author's phone and fax: +34644448558

Corresponding author's e-mail address: vladimir.baulin@urv.cat

Running title: Microtubule orientation induced by mechanical stress

**Keywords: plant cells | cytoskeleton | cortex | gravitropism | phyllotaxis**

ABSTRACT

Directional growth caused by gravitropism and corresponding bending of plant cells has been explored since 19th century, however, many aspects of mechanisms underlying the perception of gravity at the molecular level are still not well known. Perception of gravity in root and shoot gravitropisms is usually attributed to gravisensitive cells, called statocytes, which exploit sedimentation of macroscopic and heavy organelles, amyloplasts, to sense the direction of gravity. Gravity stimulus is then transduced into distal elongation zone, which is several mm far from statocytes, where it causes stretching. It is suggested that gravity stimulus is conveyed by gradients in auxin flux. We propose a theoretical model that may explain how concentration gradients and/or stretching may indirectly affect the global orientation of cortical microtubules, attached to the cell membrane and induce their dynamic reorientation perpendicular to the gradients. In turn, oriented microtubules arrays direct the growth and orientation of cellulose microfibrils, forming part of the cell external skeleton and determine the shape of the cell. Reorientation of microtubules is also observed in reaction to light in phototropism and mechanical bending, thus suggesting universality of the proposed mechanism.

INTRODUCTION

Main ingredients of external skeleton in plant cells are cellulose microfibrils (1). They help to maintain the shape of cells and prevent plant cells from bursting due to high internal turgor pressure (1,2). Cellulose microfibrils have high tensile strength and provide the cell wall mechanical strength and stiffness (2). Since cellulose arrays are naturally anisotropic, the anisotropy of the cell growth is controlled by the orientation of the cellulose microfibrils arrays (3). In turn, the deposition of cellulose microfibrils is directed by highly aligned microtubule (MT) arrays (4), which serve as template for directed growth of cellulose microfibrils (4).

MTs are hollow cylinders of 25 nm external and 15 nm internal diameter (1) and the length ranged from few nm to several microns. They are highly dynamic by their nature (5), and MT ends are constantly switching between polymerization and depolymerization, thus making the length of MTs intermittent. This property of MTs is called dynamic instability (5–7); a plus-end of a MT is usually more dynamic than a minus-end (7–9).

Cortical MTs in plant cells are organized in parallel arrays adjacent to the cell membrane (10). There is an evidence (11,12) that cellulose microfibrils may be deposed in the same direction as cortical MTs during the plant growth. Depolymerizing drugs, ethylene and various agents (11,13,14) affecting orientation of MTs, also change the orientation of cellulose microfibrils arrays (15). Aligned cellulose microfibrils provide anisotropy of the cell wall, which is more rigid in the direction parallel to cellulose arrays than in perpendicular direction. This anisotropy allows to transduce the isotropic turgor pressure into a directional cell growth (11,16–18). However, it is noteworthy that cellulose microfibril orientation in several cases was not affected by the disruption of cortical microtubules (19). Thus, anisotropy in plant cell wall rigidity irrevocably fixes the direction of growth of the cell leading to irreversibility of the processes such as cell division and cell elongation (20,21). We focus in particular on gravitropism and phyllotaxis, although the biochemical background is similar to other types of tropisms, such as phototropism or chemotropism (22).

Cell constituents, molecules and their aggregates, are too small to sense the gravitational field directly, however roots of many plants are able to sense the direction of gravity with the help of statocytes, specific cells located in the growing tip of roots or shoots (23). Statocytes can efficiently perceive the direction of the gravity and direct plant growth along the gravity vector. Directed growth is observed only in the presence of gravitational or centrifugal force and disappear in the absence of gravity or when the direction of gravity is altered (2,24). The perception of gravity in statocytes is usually attributed to amyloplasts, macroscopic, heavy organelles that sediment in a lower part of the cell (23–25) in root gravitropism and can exhibit saltatory upward movements in shoot gravitropism (3,25).

Sedimenting amyloplasts inside statocytes are probably the main driving force for root gravitropism. Since the radius of amyloplasts, $r$, is about few microns (26,27), their concentration corrected for buoyancy is $\sim\!\Delta\rho=0.5$ $g/ml$ (28) and $g\approx 9.8$ $m/s^2$, the resulting sedimentation force of one amyloplast at the bottom of the cell is $\Delta\rho g(4/3)\pi r^3\sim\!1$ $pN$ (29). Such force is insufficient to significantly stretch the membrane of statocyte and elongate it. However, the elongation growth occurs in epidermis, where MTs reorientation takes place. The distance between elongation zone and root cap is several mm, or ~ 20 cell layers (25,30,31). Amyloplast sedimentation causes alternation of auxin flux, which causes reorientation of microtubules(32–34) supported by other signals, such as $Ca^{2+}$ ions (35). Most probably, gradients in hormonal fluxes change the stability of MTs depending on their direction (36–38) (Consult Figure 1).

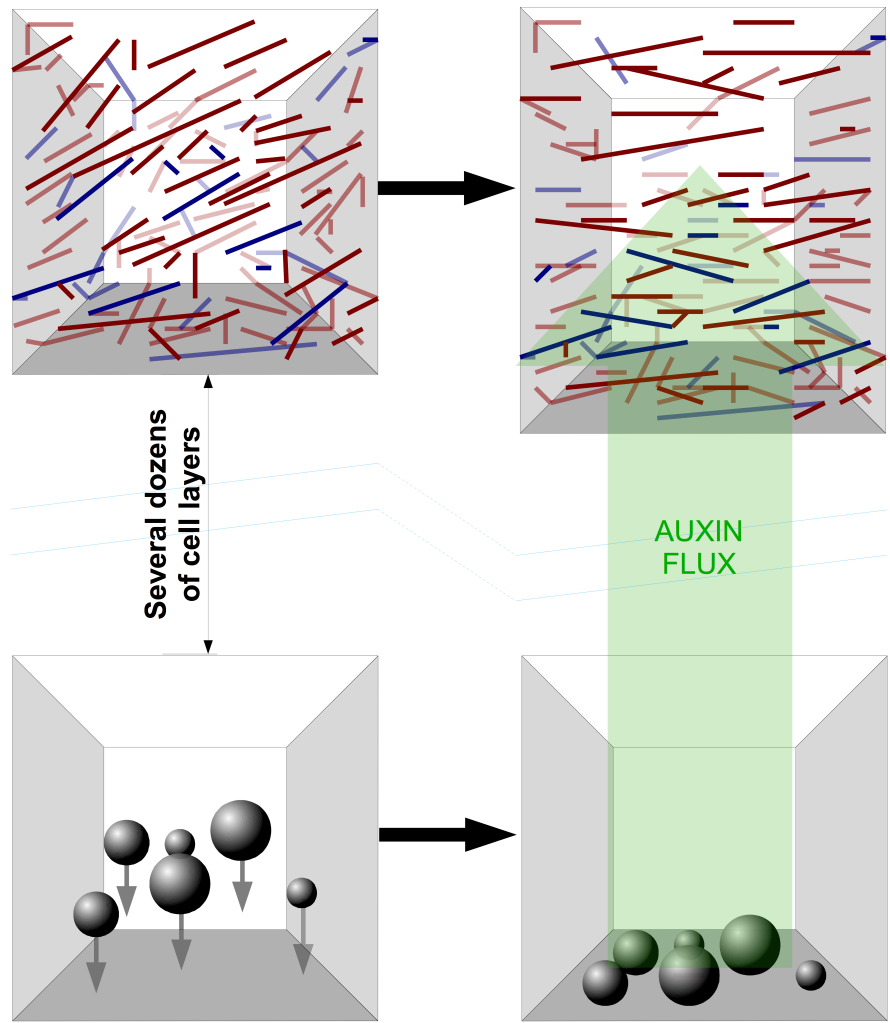

**Figure 1.** Cell elongation caused by gravity-induced sedimentation of amyloplasts. Cell stretching induces reorientation of MTs perpendicular to elongation direction.

In turn, there is an evidence that MTs respond directly to mechanical stress of the cell (2,39). It may be observed during phyllotaxis in shoot apical meristem (17,21,40). This suggests that MTs might feel mechanical stress and reorient themselves inducing anisotropic cellulose deposition. Such mechanism may be also observed in the experiments with shoot apical meristem compression, laser ablations of its cells or weakening of the cell wall (17,41,42).

Here we provide a theoretical basis for MT reorientation caused by concentration gradients of chemical agents or mechanical cell elongation. Our model is based on the mechanism of collective self-orientation of cortical MTs induced by mutual collisions and re-growing of individual MTs. This mechanism was first proposed in (43). It was shown that collisions between individual MTs may spontaneously lead to orientation domains with highly aligned MTs from initially disordered array. This simple model assumes a MT as a rigid rod that can grow at a plus-end and shorten at a minus-end, while the rate of growth at a plus-end is altered by the collisions with other MTs. It was shown that the anisotropy in the rates of growth at a plus-end due to collisions is enough to induce collective phenomena of MT self-ordering into aligned domains with preferential orientation. The selection of preferential orientation in the domains is similar to evolution selection, where MTs with "incorrect" orientation disassemble and disappear, leaving space to "correctly" aligned and thus, longer and older MTs (43).

This minimal model based on age and length discrimination was further extended and improved in consequent theoretical models (44–48). The models consider two dimensional (2D) movements because cortical MTs in plant cells are attached to plasma membrane forming a 2D array (1,2). Since the movements in 2D are much more restricted than in 3D, the probability of collisions is high even for relatively low concentrations. Direct observation of collisions between MTs in the cortex array and measure of collision rates (49) have shown that the probabilities of catastrophes and consequent shrinkage of MTs, rescue and continuation of growth depend on angles of collisions. Shallow angles favor continuation of growth, while perpendicular collisions may provoke catastrophe and disassembling of the MT (49). The models based on these results (44,45) predict the self-orientation induced by collisions and show that zippering between MTs with similar orientation play an important role in onset of ordering. Another model of collision induced reorientation of MTs (47) focuses on phase transitions in MT arrays and predicts the existence of three phases: isotropic phase, weakly ordered nematic phase and highly ordered nematic phase. A similar model (46,48) explicitly includes dynamic instability of a plus-end and uses more realistic parameters for growth and shrinkage rates of MT ends. In addition, this model implements a set of rules driving MTs in case of their collisions: induced catastrophe, plus-end entrainment (zippering) and intersection are possible depending on the angle between colliding MTs (46,48). This model predicts orientation induced by collisions and competition between domains with different orientations.

MATERIALS AND METHODS

Here we use a modified model of collision induced ordering in 2D MTs arrays that incorporates essential parts of the previous models. We model MTs as rigid rods which can switch between shrinking and growing at their plus-end (7,50). Switching between polymerization and depolymerization on a plus-end happens according to preset catastrophe and rescue rates. In case of collision with another MT the plus-end of a MT stops growing (51); however, it may still experience catastrophe with a preset rate and start shrinking (7). This model is summarized at Figure 2 and the parameters (Table 1) describing individual MT dynamics correspond to other models and to experimental data. The parameters in Table 1 are picked after several trials and to correspond the parameters used by other authors (44–48). Each MT is characterized by its length, position and orientation. We consider that orientation is set while MT is nucleated and does not change during its lifetime. Although MTs are attached to the membrane, which is well-known to be fluid, their weight is large enough to prevent them from diffusion or rotation caused by membrane movements (5). The only way to reorient an array of MTs is to eliminate it by complete disassembly the MTs with "incorrect" orientation and inject new MTs with "correct" orientations. Zippering effect for shallow contact angles (44,45,49) would only enhance the suggested mechanism and lead to faster ordering.

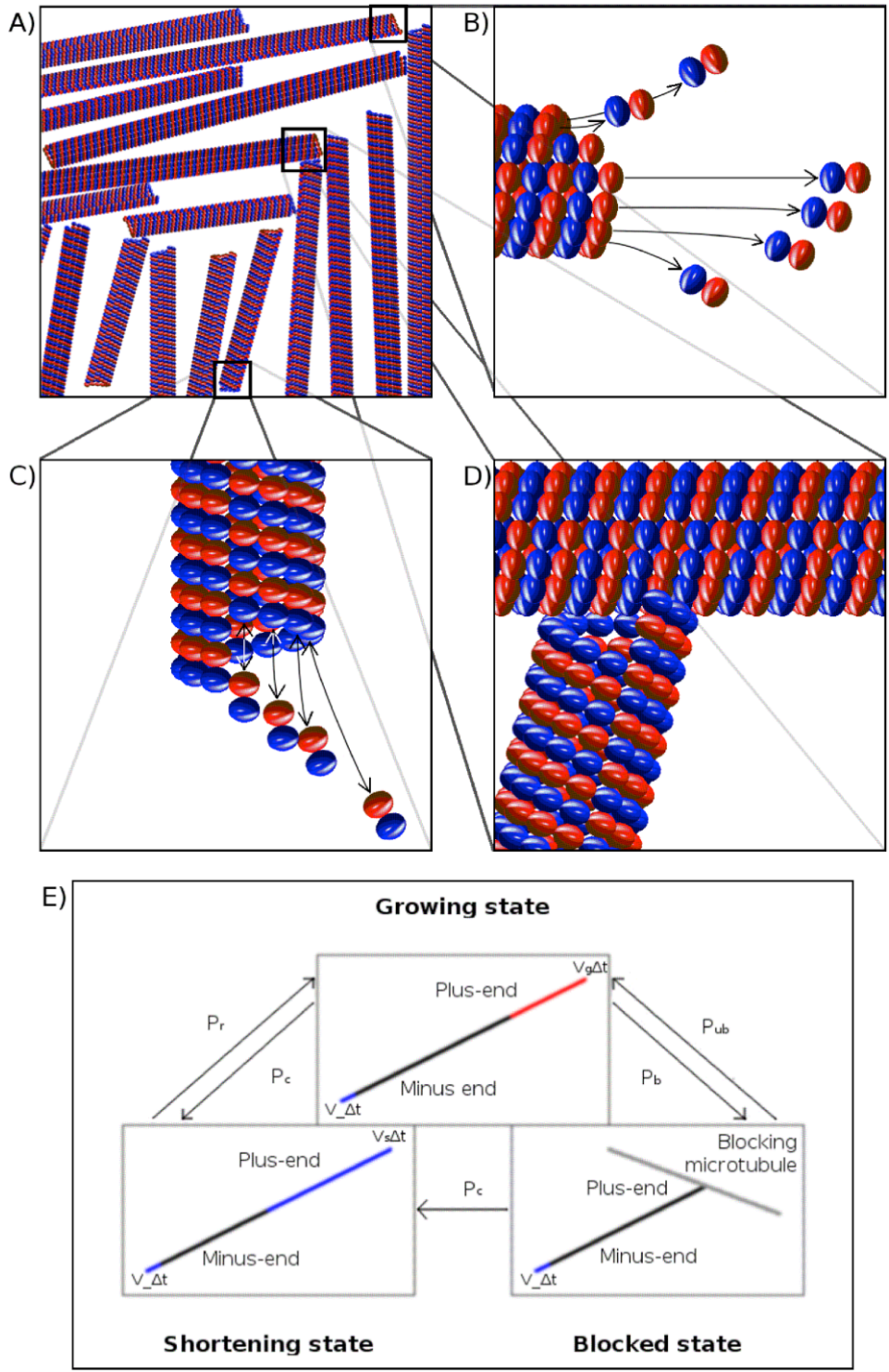

**Figure 2.** A) Schematic representation of interacting MTs in cell cortex. Schematic picture of
B) a minus-end disassembly; C) a plus-end assembly and disassembly; D) a growing MT blocked by
another one. E) The model: $P_c$ and $P_r$ are the catastrophe and rescue rates, $P_b$ and $P_{ub}$ are probabilities of blocking and unblocking respectively, $v_g$, $v_s$ and $v_-$ are the speeds of plus-end growing and shortening and minus-end shortening relatively.

| Parameter | Disordered state | Ordered state |
|---|---|---|
| Plus-end growth rate $v_g$, *nm/s* | 70 | 80 |
| Plus-end shortening rate $v_s$, *nm/s* | 225 | 200 |
| Minus-end shortening rate $v_-$, *nm/s* | 15 | 40 |
| Catastrophe probability $P_c$, *1/s* | 64x10$^{-3}$ | |
| Rescue probability $P_r$, *1/s* | 124x10$^{-3}$ | |
| Nucleation rate $P_n$, *1/s* | 5000 | |

**Table 1.** The parameters used in the model. These parameters are based on experimental data by (49,50)].

According to the minimal model described in (43), MT is a rigid rod that can grow at a plus-end and shorten at a minus-end. Collisions with other MTs perturb the growth of MTs, which itself is sufficient to induce a global order in the system without even excluded volume effects that are necessary for ordering in ordinary lyotropic liquid crystals, which comes from purely collective and kinetic interaction between MTs. However, this model, due to its simplicity, assumes infinite and unrestricted growth of perfectly aligned MTs, thus the model should include dynamic instability of the plus-end, which would lead to a stationary state of ordered MTs (46–48).

Thus, in contrast to (43), our model is a three-state model, where MT may exist in the growing state (*g*), shrinking state (*s*) and blocked state (*b*), which is similar to (46,48). The length and the position of each MT change with time. Every time interval $\Delta t$ the minus-end of each MT is shortened by $v_-\Delta t$, where $v_-$ is the speed of shrinkage of the minus-end. The speed of elongation of the plus-end is $v_g$ and the speed of its shrinkage is $v_s$. The plus-end of initially growing MT can experience catastrophes with the rate $P_c$ and rescues with the rate $P_r$; can be blocked due to collisions with the rate $P_b$ and unblocked with the rate $P_{ub}$, thus providing stochastic oscillations of MT length. If the length of a MT goes to 0, it disappears, while new MTs are created with the nucleation rate $P_n$ (Figure 2). The balance between growing and shrinking, nucleation and disappearing of MTs in a steady state insures dynamic stability of the average number of MTs and their average length. The values of parameters are given in Table 1. If a MT occasionally intersects any other MT, the growth is rejected and the MT passes from growing to blocked state $b$. The shrinkage and growth can be reversed stochastically according to the given probabilities. Thus, every time step the length of a MT either increases (state $g$) by $(v_g\text{-}v\text{-})\ \Delta t$ or decreases (state $s$) by $(v_s\text{+}v\text{-})\ \Delta t$ or decreases (state $b$) by $v\text{-}$. $\Delta t$ (Figure 2). The model by (44) includes complex collision rules, requiring the introduction of a critical entrainment angle $\Theta_z$. They also considered cell edge effects. Both mechanisms can significantly improve the orientation of MTs. Nevertheless, we do not take them into account as we believe that these two mentioned effects would only promote the orientation.

We consider that the seeds of new MTs are nucleated homogeneously and with random angles. MTs cannot change their orientation, but they can change their states between growing ($g$), shortening ($s$) and blocked ($b$). Introducing the corresponding surface concentrations, $c_g$, $c_s$ and $c_b$, the total length concentration $k(\Theta,t)$, where $\Theta$ is the direction of individual MT and $l$ is its length, can be written in the following form:

$$k(\Theta,t) = \int_0^\infty l\Big[c_g(l,\Theta,t) + c_s(l,\Theta,t) + c_b(l,\Theta,t)\Big]dl. [1]$$

A set of evolution equations are written as

$$\frac{\partial c_g}{\partial t} = -v_g \frac{\partial c_g}{\partial l} + P_r c_s - P_c c_g + v_- \frac{\partial c_g}{\partial l} + P_{ub}c_b - P_b c_g, [2]$$

$$\frac{\partial c_s}{\partial t} = v_s \frac{\partial c_s}{\partial l} - P_r c_s + P_c c_g + P_c c_b + v_- \frac{\partial c_s}{\partial l}, [3]$$

$$\frac{\partial c_b}{\partial t} = v_- \frac{\partial c_b}{\partial l} - P_{ub}c_b + P_b c_g - P_c c_g, [4]$$

where $P_r c_s = \Phi_r[c_s]$ and $P_c c_g = \Phi_c[c_g]$ along with $P_c c_b = \Phi_c[c_b]$ are the spontaneous flux terms (46) responsible for rescue and catastrophe respectively; $-v_g(\partial c_g/\partial l) = \Phi_g[c_g]$, $vs(\partial c_s/\partial l) = \Phi_s[c_s]$ and $v_-(\partial c_i/\partial l) = \Phi_-[c_i]$ $(i=\{s,g,b\})$ describe growth and shrinkage of plus-end and shortening of minus-end respectively (6,43,46,48). Blocking term is determined by the collisions between MTs when a growing MT collides with another MT with a rate $P_b c_g(\Theta) = v_g c_g(\Theta) \int d\Theta' \sin|\Theta - \Theta'| k(\Theta') = \Phi_b[c_g, k]$, where $\sin|\Theta - \Theta'|$ defines the cross section of collisions. This term reminds the second virial coefficient of the Onsager theory (52), but has a completely different physical origin. The unblocking term $P_{ub}(c_i, k)c_b$ is connected with the possibility of a previously blocked plus-end of a MT to restart its growing due to disassembly of a blocking MT, no matter if it is a shrinkage of a plus-end or shortening of a minus-end. It is related to the concentration $c_b(l,t,\Theta)$ as following $P_{ub}(c_i, k)c_b(\Theta) = \Phi_{ub}[c_b, c_i, k]$.

The initial conditions are the following: at $t=0$ the concentration of shortening and blocked MTs are equal zero $c_s(l,0,\Theta)=0$, $c_b(l,0,\Theta)=0$ while $c_g$ is connected with the nucleation rate $P_n$ as $(v_g-v_s)c_g(0,t,\Theta) = P_n(\Theta)/2\pi$, $c_g(l>0,0,\Theta)=0$.

The first model (43) lacked dynamic instability, i.e. missed $\Phi_c$ and $\Phi_r$ terms in eqs (2)-(4), which are responsible for stabilization of the length of MTs in ordered arrays and thus these terms are essential for the stationary state in ordered arrays. The model by Hawkins *et al.* (2010) did not have the minus-end disassembly, i.e. terms $\Phi_-$, $\Phi_b$ and $\Phi_{ub}$, but it included $\Phi_{inducedcat.}$, $\Phi_{zipper}$ and $\Phi_{reactivation}$. It is possible to relate the population of "inactive" segments with the population of blocked segments. Note, that "zippering" in this model is similar to "blocked" described by the present model, where the reactivation occurs due to disassembly of blocking MTs, in contrast to disassembly of an active segment of the MT in zippering event. The argument that microtubules undergo "zippering" may lead to potential interactions with other microtubules that differ significantly from those blocked MTs, which may not be practical in Monte Carlo model. We believe that addition of extra interactions or changing the nature of interaction will not change the basic physics of the phenomenon.

In order to get exact theoretical description one has to solve reaction-advection equations 2-4. However, these equations are interconnected since the blocking and unblocking coefficients are not constant. Steady state solution (no dependence on $t$) solutions for $c_g$ and $c_s$ are given by

$$c_g(l,\Theta) = A(\Theta)e^{-l/\bar{l}}, [5]$$
$$c_b(l,\Theta) = B(\Theta)e^{-l/\bar{l}}. [6]$$

Showing exponential length distributions with average length in the direction $\Theta$: $\bar{l} = -g^{-1}$.

RESULTS

We performed two types of simulations: microtubule orientation in response to concentration gradients and mechanical stress of the cortex. Both stimuli may induce oriented self-organization of MTs.

**Orientation of MTs induced by concentration gradients**

Chemical agents, e.g. hormone of growth auxin or ethylene, are often reported to modify MT dynamics (33). The mechanism of such influence is not well understood. It is suggested (37) that in animal cells chemical agents catalyze or inhibit the action of microtubule-associated proteins (MAPs). With certain reservations, it can also be the case of plant cells. These proteins change dynamic parameters of MTs, and generally affect the growth velocities or catastrophe rates (37). Concentration gradients thus may affect global dynamics of MTs.

We assume that chemical agents influence dynamic instability parameters, such as catastrophe rate, which becomes higher in the direction of the concentration gradient. This assumption is consistent with experimental data showing that during gravitropic response polarity in distribution of auxin transporters, proteins of PIN family, is generated (41). These proteins are linked with MT orientation regulator CLASP (53–55) and with microtubule-associated protein MAP65 (56). There is also an evidence that hormones act through regulation of microtubule severing, which requires the action of GTPases (57); the only dynamic event which is connected with GTP conversion to GDP is catastrophe (5–7).

Two sets of parameters are selected to induce spontaneous organization of MTs (Table 1). An increase in catastrophe rate in one direction breaks the symmetry between MTs oriented along the gradient and against it. This asymmetry may induce global order of MT arrays in one direction. To investigate this phenomenon in detail we change the gradient of catastrophe rate. We assume that after 1000 timesteps the array is already stable, thus the significance of this change can be tested easily. This change indeed leads to reorientation of already formed MT arrays, as shown in Figure 2.

The degree of MT orientation can be described by the nematic order parameter, $\sigma(\Omega) = \overline{\cos^2(\Theta - \Omega)}$ where $\Omega$ is the direction of the director and the bar signifies the ensemble average. However, MTs has intermittent length and the ordering of the domains is determined by long MTs (43). Thus, the order parameter need to include the length and we use the following function (43)

$$\sigma_l(\Omega) = \overline{l^2 \cos^2(\Theta - \Omega)}.$$

With this, an anisotropy ratio $S_l$ can be defined as (43):

$$S_l = \frac{\sigma(\Omega_{max}) - \sigma(\Omega_{min})}{\sigma(\Omega_{max}) + \sigma(\Omega_{min})}$$

and the dominant angle is given by (43):

$$\tan(2\Omega_l) = \frac{\overline{l^2 \sin^2 2\Theta}}{\overline{l^2 \cos^2 2\Theta}}.$$

First set of parameters lead to slower reorientation: although sample array's anisotropy ratio soon recovers its high value (See Figure 3F)), it takes more time for MTs to reorient, as depicted by sample array's dominant angle in Figure 3E). Eventually, the system reaches the state shown on Figure 3B), which takes around 10000 steps. Most probably, this is due to existence of domains during reorientation, while reorientation of MTs described by second set of parameters occurs with the complete disruption of microtubules (Figure 3C-D)). Thus, the difference in reorientation behavior is primarily controlled by the populations of blocked (*b*) and growing (*g*) MTs, while shortening MTs (*s*) shrink with the same rate $v_s+v_-=240$ *nm/s*.

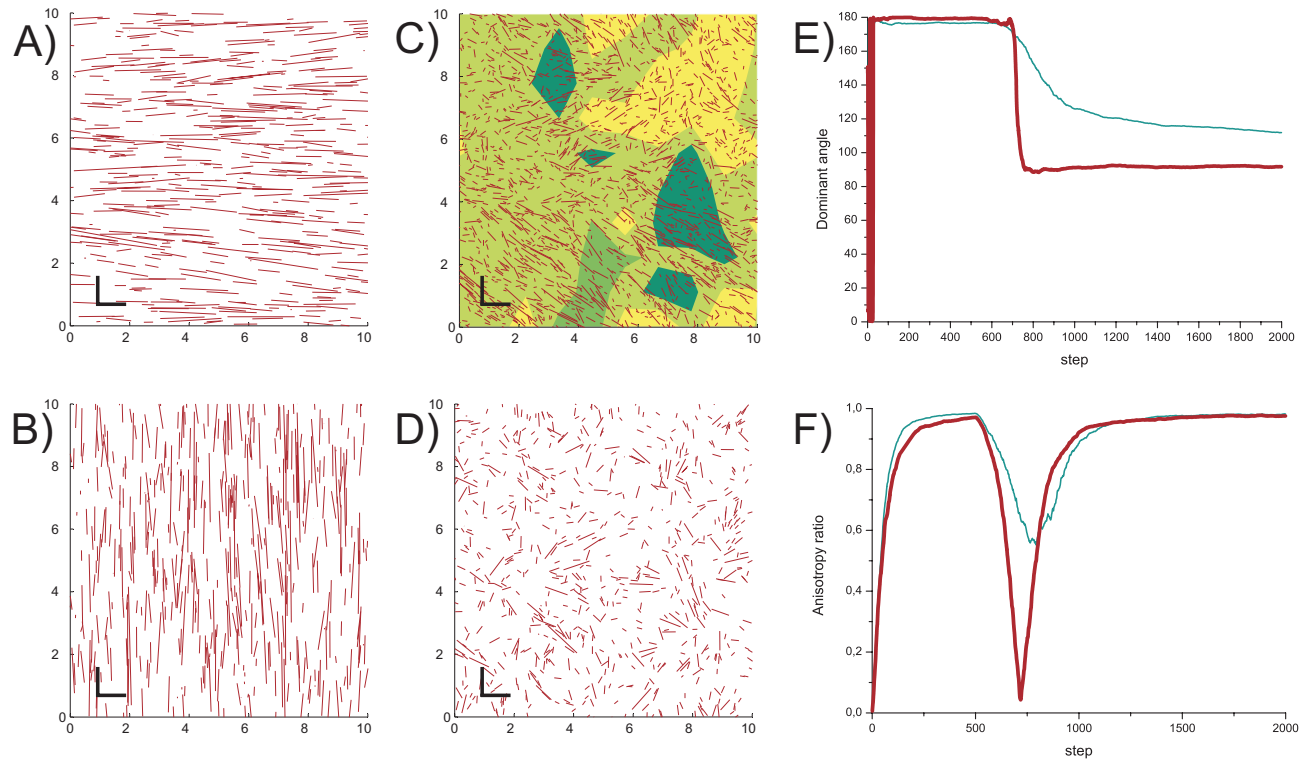

**Figure 3.** MTs A) before and B) after C-D) during reorientation for C) first and D) second set of parameters in Table 1. Only 1000 longest MTs are shown. Bar indicates 1 μm. Anisotropy ratio on snapshot C) is reproduced by the scale of green. For two sets of parameters in Table 1: E) dominant angle versus simulation timestep; F) anisotropy ratio versus simulation timestep. First set of parameters is represented by grey line, second - by cyan line.

### Compression-induced orientation in disordered arrays

Cortex MTs attached to the cell membrane may form stationary states of oriented domains or stay in disordered state depending on the rates of growth and shrinkage, catastrophe rates and concentration (46–48). Mechanical stretching or compression of the cell membrane anisotropically changes the distance between the MTs, thus affecting the concentration in the direction of applied force. This, in turn, may induce orientation in disordered arrays or reorient MTs in ordered arrays. In the following, we investigate the effect of stretching and compression on these two initial stationary states, (i) isotropically oriented, disordered array of MTs and (ii) oriented domains of MTs are then subject to stretching and compression.

Set of parameters corresponding to initially disordered arrays (Table 1) leads to formation of stationary disordered state, characterized by the balance between growing, shrinking and blocked MTs. Thus, stationary average total number of MTs is provided by the balance between MTs that disappear due to collisions with other MTs or due to catastrophes events, and new-born MTs appearing with random directions. Freshly appearing MTs correspond to nucleation sites that are fixed in the cortex, thus we assume that the total number of new-born MTs per time step and per area is kept constant. However, stretching or compression change the distance between nucleation sites and thus, the nucleation rate, being inversely proportional to the area, may vary with the direction.

Stationary disordered array of *10x10 μm²* (Figure 4A)) is compressed in one direction by two times (Figure 4B)). This compression provokes orientation of MTs in the direction of compression. In contrast, the stretching by two times of the array does not lead to orientation. This goes inline with the observations of MT orientation perpendicular to the gravity vector in gravitropism in roots (32) and may be related to coordinated patterns of MT arrays governed by mechanical stress (18).

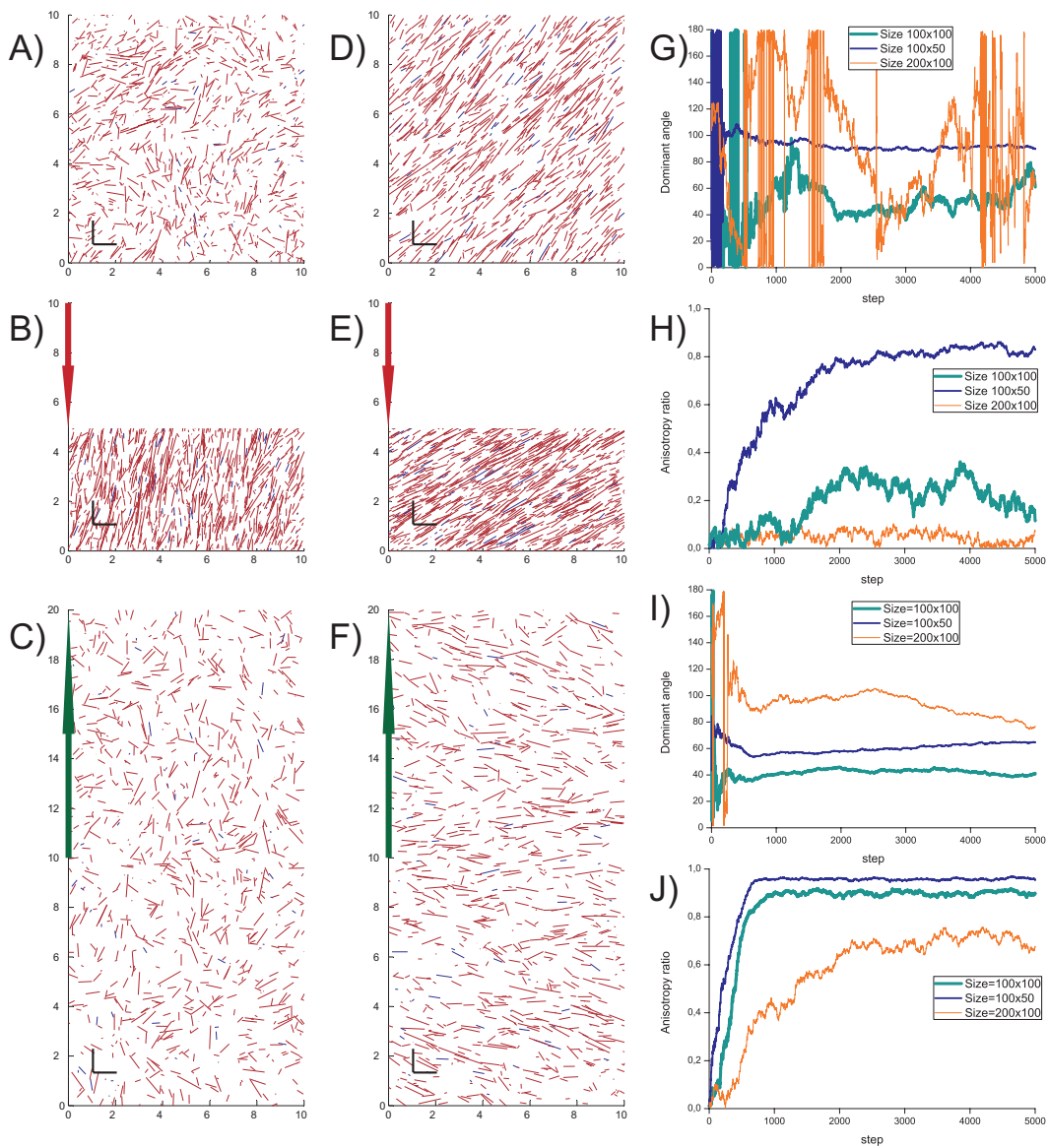

**Figure 4.** Snapshots of MT arrays (only 1000 longest MTs are shown) evolving from A) initially homogeneous and isotropic array 10x10 μm under B) compression and C) stretching. Shrinking MTs (state *s*) are shown in blue, other MTs (state *b* and *g*) are shown in red. Bar indicates 1μm. D-F) Snapshots of MT arrays evolving D) from ordered array 10x10 μm under E) compression and F) stretching. Dominant angle versus simulation timestep G) for initially isotropic array in A-C) and I) oriented array in D-F). Anisotropy ratio versus simulation timestep H) for initially isotropic array in A-C) and J) oriented array in D-F).

The anisotropy ratio $S_l$ and the dominant angle $\Theta_l$ as function of time are given in Figure 4G) and H).

### Stretching-induced disorientation in ordered arrays

Once a stationary ordered array is formed (Figure 4D)), the orientation of the array is governed by longer MTs (43), as they are more stable, and thus having biggest life expectancy and hence more persistent. Since the angles of MTs in this model are irrevocably fixed during the life cycle, changing the direction of the ordered array implies disassemble of these leading MTs, which may require collisions with domains with even longer MTs. Thus, compression of the ordered arrays only decreases the distance between MTs that may increase the order, but may not lead to disassemble of MTs. This is shown in Figure 4E) and the corresponding plot of the dominant angle and anisotropy ratio $S_l$, Figures 4I-J).

In turn, stretching of the cortex increase the distance between MTs, thus new-born MTs may grow longer before they collide with the dominant MTs and bring more random orientations into arrays. Figures 4F) and I-J) show that stretching may considerably reduce the order in the arrays, and, in principle, may lead to complete disorientation of the arrays.

Both effects, disorientation of the arrays due to stretching and orientation induced by compression, may work together to help to reorient MT arrays in response to mechanical stress on the cortex.

DISCUSSION

Our results show that microtubules may reorient in response to inhomogeneous change of catastrophe rate (which might be caused by gravitropism due to the flux of auxin) or in response to the change of cell geometry which might be caused by phyllotaxis due to extension and/or compression of the cell. It is important to notice that the idea of compression is mostly artificial and such processes are rarely observed *in vivo*; however, some *in vitro* experiments support our results that microtubules would be oriented orthogonally to the cell axis (41,42). *In vivo* compression of the cell (if it exists) shall be gradual and not sudden, as it occurs in our simulations. However, our model allows us to perform more detailed simulations of a constant compression of the cell after some modifications, which is subject of further research.

Experimental data show that three possible outcomes might occur with colliding microtubule during collision: induced catastrophe, crossover and "zippering". First two turnovers do not lead to any flux which is not present in our model; they can only change the possibilities of the present fluxes (See Figure 4E)). As for the "zippering", its flux can be written as $\Phi_{zipper}[c_g, k] = v_g c_g(\Theta) \int d\Theta' P_{zipper} \sin|\Theta - \Theta'| k(\Theta')$ (46), which corresponds to a "blocking" flux in our model. The reverse "reactivation" flux would also correspond to "unblocking" flux (46). Thus, our model only accounts for blocking for the sake of simplicity. In turn, the edge effects were reported to alter the orientation of microtubules only slightly (44,45) as compared to stretching in phyllotaxis or auxin flux in gravitropism.

The model of kinetic orientation of MT arrays due to collisions was used to study possible effects of concentration gradients of hormones and mechanical deformation (compression and stretching) of the cortex in plant cells. In gravitropic response, concentration gradients of auxin influence catastrophe and growth rates of MTs, which in turn may lead to reorientation of MT arrays transverse to the direction of the gradient. In phyllotactic arrangement of leafs on stem, the mechanical stretching of initially disordered MT arrays itself may induce a global order, where MTs are oriented along the action of mechanical stress. Ordered MT arrays, in turn, can direct cellulose microfibrils assembly, thus resulting in oriented microfibrils growth. Orientation of cellulose microfibrils in a certain direction consequently leads to anisotropy of the cell wall, which, in turn, combined with turgor pressure indicates the direction of cell growth.

In future work, a more detailed model including different types of interactions between microtubules on atomistic level may provide a description of the effects of slow continuous extension and/or compression.